\documentstyle[12pt,epsfig,psfig,amsmath]{article}

\begin{document}
\title{Spinor Casimir Densities for a Spherical Shell in the Global Monopole
Spacetime}
\author{A. A. Saharian$^{1,2}${\thanks{E-mail:
saharyan@www.physdep.r.am}}\,
 and E. R. Bezerra de Mello$^{3}${\thanks{E-mail: emello@fisica.ufpb.br}}\\
\textit{$^1$Department of Physics, Yerevan State University,}\\
\textit{ 375049 Yerevan, Armenia}\\
\textit{$^2$the Abdus Salam International Centre for Therietical
Physics,}\\
\textit{ 34014 Trieste, Italy, } \\
\textit{$^3$Departamento de F\'{\i}sica-CCEN, Universidade Federal da Para\'{\i}ba}\\
\textit{58.059-970, J. Pessoa, PB C. Postal 5.008, Brazil}}
\maketitle
\begin{abstract}
We investigate the vacuum expectation values of the
energy-momentum tensor and the fermionic condensate associated
with a massive spinor field obeying the MIT bag boundary condition
on a spherical shell in the global monopole spacetime. In order to
do that it was used the generalized Abel-Plana summation formula.
As we shall see, this procedure allows to extract from the vacuum
expectation values the contribution coming from to the unbounded
spacetime and explicitly to present the boundary induced parts. As
to the boundary induced contribution, two distinct situations are
examined: the vacuum average effect inside and outside the
spherical shell. The asymptotic behavior of the vacuum densities
is investigated near the sphere center and surface, and at large
distances from the sphere. In the limit of strong gravitational
field corresponding to small values of the parameter describing
the solid angle deficit in global monopole geometry, the
sphere-induced expectation values are exponentially suppressed. As
a special case we discuss the fermionic vacuum densities for the
spherical shell on background of the Minkowski spacetime. Previous
approaches to this problem within the framework of the QCD bag
models have been global and our calculation is a local extension
of these contributions.
\end{abstract}

\bigskip

PACS number(s): 03.70.+k, 04.62.+v, 12.39.Ba

\bigskip

\section{Introduction}
Different types of topological defects \cite{V-S} may have been
formed during the phase transitions in the early universe.
Depending on the topology of the vacuum manifold ${\cal M}$ these
are domain walls, strings, monopoles and textures corresponding to
the homotopy groups $\pi_0({\cal M})$, $\pi_1({\cal M})$,
$\pi_2({\cal M})$ and $\pi_3({\cal M})$, respectively. Physically
these topological defects appear as a consequence of spontaneous
breakdown of local or global gauge symmetries of the system
composed by self-coupling scalar Higgs or Goldstone fields,
respectively. Global monopoles are spherically symmetric
topological defects created due to phase transition when a global
symmetry is spontaneously broken and they have important role in
the cosmology and astrophysics.

The simplest theoretical model which provides global monopoles has been proposed
a few years ago by Barriola and Vilenkin \cite{B-V}. This model is composed by a
self-coupling iso-scalar Goldstone field triplet $\phi^a$, whose original global
$O(3)$ symmetry is spontaneously broken to $U(1)$. The matter field plays
the role of an order parameter which outside the monopole's core acquires a
non-vanishing value. The main part of the monopole's energy is concentrated into its
small core. Coupling this system with the Einstein equations, a spherically
symmetric metric tensor is found. Neglecting the small size of the monopole's core,
this tensor can be approximately given by the line element
\begin{equation}
ds^{2}=dt^{2}-dr^{2}-\alpha ^{2}r^{2}\left( d\theta ^{2}+\sin ^{2}\theta
d\phi ^{2}\right) \ ,
\label{mmetric}
\end{equation}
where the parameter $\alpha^2$, smaller than unity, depends on the
symmetry breaking energy scale and codifies the presence of the
global monopole\footnote{In fact the parameter $\alpha^2=1-8\pi
G\eta^2$, being $\eta$ the energy scale where the global symmetry
is spontaneously broken.}. This spacetime corresponds to an
idealized point-like global monopole. It is not flat: the scalar
curvature $R= 2(1-\alpha^2)/r^2$, and the solid angle of a sphere
of unit radius is $\Omega= 4\pi\alpha^2$, so smaller than the
ordinary one. The energy-momentum tensor associated with this
object has a diagonal form and its non-vanishing components read
$T^0_0=T^r_r=(\alpha^2-1)/r^2$.

The quantum effects due to the point-like global monopole
spacetime on the matter fields have been considered in Refs.
\cite{M-L} and \cite{EVN} to massless scalar and fermionic fields,
respectively. In order to do that, the scalar and spinor Green
functions in this background have been obtained. More recently the
effect of the temperature on these polarization effects has been
analysed in \cite{C-E} for scalar and fermionic fields. The
calculation of quantum effects on massless scalar field in a
higher dimensional global monopole spacetime has also been
developed in \cite{E}.

Although the deficit solid angle and also curvature associated
with this manifold produce non-vanishing vacuum polarization
effects on matter fields, the influence of boundary conditions
obeyed by the matter fields on the vacuum polarization effects
have been investigated. The Casimir energy associated with massive
scalar field inside a spherical region in the global monopole
background have been analyzed in Refs. \cite{MKS,EVN1} using the
zeta function regularization method. More recently the Casimir
densities induced by a single and two concentric spherical shells
have been calculated \cite{A-M,Saha03b} to higher dimensional
global monopole spacetime by making use of the generalized
Abel-Plana summation formula \cite{Saharian,Saha00}. This
procedure allows to develop the summation over all discrete mode.
Here we shall calculate the Casimir densities for fermionic fields
obeying MIT bag boundary condition on the spherical shell in the
point-like global monopole spacetime. Specifically we shall
calculate the renormalized vacuum expectation values of the
energy-momentum tensor and the fermionic condensate in the regions
inside and outside the spherical shell. As we shall see using the
generalized Abel-Plana summation formula, all the components of
the vacuum average of the energy-momentum tensor can be separated
in two contributions: boundary dependent and independent ones. The
boundary independent contribution is similar to previous result
obtained in \cite{EVN} using different approach. It is divergent
and consequently in order to obtain a finite and well defined
expression we must apply some regularization procedure. The
boundary dependent contribution is finite at any strictly interior
or exterior point and does not contain anomalies. Consequently, it
does not require any regularization procedure. Because the
analysis of boundary independent term has been performed before,
in this present analysis we shall concentrate on the boundary
dependent part. Taking $\alpha =1$, from our results in this paper
as a special case we obtain the fermionic Casimir densities for a
spherical shell on background of the Minkowski spacetime.
Motivated by the MIT bag model in QCD, the corresponding Casimir
effect was considered in a number of papers
\cite{Bend76,Milt80,Milt81,Milt83,Baac83,Blau88,Eliz98,Cogn01}
(for reviews and additional references see
\cite{Most97,Plun86,Milt02,Bord01}). To our knowledge, the most of
the previous studies were focused on global quantities, such as
the total vacuum energy and stress on the surface. The density of
the fermionic vacuum condensate for a massless spinor field inside
the bag was investigated in Ref. \cite{Milt81} (see also
\cite{Milt02}). In the considerations of the Casimir effect it is
of physical interest  to calculate not only the total energy but
also the local characteristics of the vacuum, such as the
energy-momentum tensor and vacuum condensates. In addition to
describing the physical structure of the quantum field at a given
point, the energy-momentum tensor acts as the source of gravity in
the Einstein equations \footnote{The effects of the back-reaction
corrections on the Einstein equation due to the vacuum
polarization produced by massless scalar field in a global
monopole spacetime has been analyzed in \cite{M-L}.}. It therefore
plays an important role in modelling a self-consistent dynamics
involving the gravitational field \cite{Birr82}. For the case of
the Minkowski bulk, our calculation is a local extension of the
previous contributions on the fermionic Casimir effect for a
spherical shell.

This paper is organized as follows: In section \ref{sec:eigfunc}
we obtain the normalized eigenfuntions for a massive spinor field
on the global monopole spacetime inside a spherical shell of
finite radius. In section \ref{sec:inside}, using the generalized
Abel-Plana summation formula, we formally obtain the vacuum
expectation value of the energy-momentum tensor considering that
the fermionic field obeys the MIT bag condition on the spherical
shell. Explicit behavior for boundary dependent term is exhibited.
The section \ref{sec:outside} is devoted to the calculation of the
vacuum expectation values for the region outside the shell. In
section \ref{sec:conc} we present our concluding remarks and leave
for the Appendix some relevant calculations.

\section{{Eigenfunctions for a Spinor Field on the Global} \label{sec:eigfunc}\\
Monopole Spacetime}

The dynamics of a massive spinor field on a curved spacetime is described by the
Dirac differential equation
\begin{equation}
i\gamma ^{\mu }(\partial _{\mu }+\Gamma _{\mu })\psi -M\psi =0 \ ,
\label{Diraceq}
\end{equation}
where $\gamma^{\mu }$ are the Dirac matrices defined in such curved spacetime, and
$\Gamma_\mu$ the spin connection defined as
\begin{equation}
\Gamma_{\mu }=\frac{1}{4}\gamma_{\nu }\nabla_{\mu }\gamma^{\nu } \ ,
\label{spincon}
\end{equation}
being $\nabla_{\mu }$ the standard covariant derivative operator. Notice that,
using the usual commutation relations for the Dirac matrices, $\{\gamma^\mu, \
\gamma^\nu\}=2g^{\mu\nu}$, we can see that
\begin{equation}
\gamma ^{\mu }\Gamma _{\mu }=\frac{1}{4}\nabla _{\mu }\gamma^{\mu }+\frac{1}
{8}\gamma^{\mu }\gamma^{\nu }\left( \partial _{\mu }\gamma_{\nu
}-\partial_{\nu }\gamma_{\mu }\right) \ .  \label{rel1Gam}
\end{equation}

After this brief introduction, let us now specialize in the
spacetime associated with the point-like global monopole whose
line element is described by (\ref{mmetric}). In order to develop
such procedure we shall adopt the following representation for the
Dirac matrices
\begin{equation}
\gamma^{0}=\left(
\begin{array}{cc}
1&0 \\
0&-1
\end{array} \right) \ ,
\gamma^{k}=\left(
\begin{array}{cc}
0&\sigma^{k} \\
-\sigma^{k}&0
\end{array} \right) \ ,
\end{equation}
given in terms of the curved space Pauli $2\times 2$ matrices
$\sigma ^k$:
\begin{eqnarray}
\sigma ^{1}&=&\left(
\begin{array}{cc}
\cos \theta  & e^{-i\phi }\sin \theta  \\
e^{i\phi }\sin \theta  & -\cos \theta
\end{array}
\right) \ ,\quad \sigma ^{2}=\frac{1}{\alpha r}\left(
\begin{array}{cc}
-\sin \theta  & e^{-i\phi }\cos \theta  \\
e^{i\phi }\cos \theta  & \sin \theta
\end{array}
\right) \ , \nonumber \\
\sigma ^{3}&=&\frac{i}{\alpha r\sin \theta }\left(
\begin{array}{cc}
0 & -e^{-i\phi } \\
e^{i\phi } & 0
\end{array}
\right) .  \label{Pauli}
\end{eqnarray}
These matrices satisfy the relation
\begin{equation}
\sigma ^{l}\sigma ^{k}=\gamma ^{lk}+i\frac{\varepsilon ^{lkm}}{\sqrt{\gamma }}
\gamma _{mp}\sigma^{p} \ ,
\label{Pauli1}
\end{equation}
where $\gamma^{lk}=-g^{lk}$ are the spatial component of metric
tensor and $\gamma$ is the corresponding determinant. $\varepsilon
^{lkm}$ is the totally anti-symmetric symbol with
$\varepsilon^{123}=1$. Here and below the latin indices $
i,k,\cdots$ run over values $1, \ 2, \ 3$. It can be easily
checked that with these representations the Dirac matrices satisfy
the standard anticommutation relations. Substituting these
matrices into formula (\ref{rel1Gam}), we can see that
\begin{equation}
\gamma^{\mu }\Gamma _{\mu }=\frac{\alpha -1}{\alpha r}\gamma ^{1} \ .
\label{rel2Gam}
\end{equation}

Let us write the four-components spinor field $\psi$ in terms of two-components ones
as
\begin{equation}
\psi =\left(
\begin{array}{c}
\varphi  \\
\chi
\end{array}
\right) \ .
\label{spinors}
\end{equation}
Assuming the time dependence in the form $e^{-i\omega t}$, from
(\ref{Diraceq}) one finds the equations for these spinors:
\begin{subequations} \label{spinoreq}
\begin{eqnarray}
\sigma^{k}\partial_{k}\varphi+\frac{\alpha-1}{\alpha r}(\hat{n}
\cdot\vec{\sigma })\varphi&=&i(\omega +M)\chi \ ,   \\
\sigma ^{k}\partial_{k}\chi +\frac{\alpha-1}{\alpha
r}(\hat{n}\cdot \vec{\sigma })\chi&=&i(\omega -M)\varphi \ ,
\end{eqnarray}
\end{subequations}
where $\vec{\sigma }=(\sigma ^{1},\sigma ^{2},\sigma ^{2})$, and
$\hat{n}=\vec{r}/r$. The angular parts of the spinors are the
standard spinor spherical harmonics $\Omega _{jlm}$ whose explicit
form is given in Ref. \cite{Berest}:
\begin{equation}
\varphi=f(r)\Omega_{jlm},\quad \chi
=(-1)^{(1+l-l')/2}g(r)\Omega_{jl'm} \ , \label{sphharmonics}
\end{equation}
where $j$ specifies the value of the total angular momentum, and
$m$ its projection, $l=j\pm 1/2$, $l'=2j-l$. Using the formula
\begin{equation}
\Omega_{jl'm}=i^{l-l'}(\hat{n}\cdot\vec{\sigma} )\Omega_{jlm} \ ,
\label{relsphhar}
\end{equation}
it can be seen that
\begin{subequations}\label{relphixi}
\begin{eqnarray}
\sigma^{k}\partial_{k}\varphi&=&i^{l'-l}\left[f^{\prime
}(r)+\frac{
1+\kappa}{\alpha r}f(r)\right]\Omega_{jl'm}, \\
\sigma^{k}\partial_{k}\chi  &=&-i\left[ g^{\prime
}(r)+\frac{1-\kappa }{ \alpha r}g(r)\right]\Omega _{jlm} \ ,
\end{eqnarray}
\end{subequations}
where we use the notation
\begin{equation}
\label{kappa}
\kappa =\left\{
\begin{array}{cc}
-(l+1), & j=l+1/2 \\
l, & j=l-1/2
\end{array}
\right. .
\end{equation}

By taking into account these relations, from (\ref{spinoreq}) we obtain the
following set of differential equations for the radial functions
\begin{subequations}\label{radeq}
\begin{eqnarray}
f^{\prime }(r)+\frac{\alpha +\kappa }{\alpha r}f(r)-(\omega
+M)g(r) &=&0 \ , \\
g^{\prime }(r)+\frac{\alpha -\kappa }{\alpha r}g(r)+(\omega
-M)f(r) &=&0 \ .
\end{eqnarray}
\end{subequations}
They lead to the second order differential equations for the separate
functions:
\begin{subequations}\label{eqforfg}
\begin{eqnarray}
f^{\prime \prime}(r)+\frac{2}{r}f^{\prime }(r)+\left[ k^{2}-\frac{\kappa
(\kappa +\alpha )}{\alpha ^{2}r^{2}}\right] f(r) &=&0 \ ,\quad k=\sqrt{
\omega ^{2}-M^{2}} \ ,  \\
g^{\prime \prime }(r)+\frac{2}{r}g^{\prime }(r)+\left[
k^{2}-\frac{\kappa (\kappa -\alpha )}{\alpha ^{2}r^{2}}\right]
g(r) &=&0 \ ,
\end{eqnarray}
\end{subequations}
with the solutions
\begin{equation}
f(r)=A\frac{Z_{|\kappa /\alpha +1/2|}(kr)}{\sqrt{r}},\quad g(r)=B\frac{%
Z_{|\kappa /\alpha -1/2|}(kr)}{\sqrt{r}} \ ,
\label{fgZ}
\end{equation}
where $Z_{\nu }(x)$ represents the cylindrical Bessel function of
the order $\nu $. The constants $A$ and $B$ are related by
equations (\ref{radeq}):
\begin{equation}
B=\frac{\mp kA}{\omega +M},\quad {\rm for}\quad j=l\pm \frac{1}{2}
\ . \label{ABrel}
\end{equation}
As a result for a given $j$ we have two types of eigenfunctions
with different parities corresponding to $j=l\pm 1/2$. These
functions are specified by the set of quantum numbers $\beta
=(\sigma kjlm)$ and have the form
\begin{eqnarray}
\psi_{\beta } &=& \frac{Ae^{-i\omega t}}{\sqrt{r}}
\left(\begin{array}{c}
Z_{\nu _{\sigma }}(kr)\Omega_{jlm} \\
in_{\sigma }Z_{\nu _{\sigma }+n_{\sigma
}}(kr)\frac{k(\hat{n}\cdot\vec{\sigma})} {\omega +M}\Omega_{jlm}
\end{array}
\right) , \label{eigfunc} \\
l & = & j- \frac{n_{\sigma}}{2},\quad \omega =\pm E, \quad
E=\sqrt{k^{2}+M^{2}} \ ,
\end{eqnarray}
where $j=1/2,3/2, \ldots $, and $m=-j, \ldots ,j$, \label{ljE}
\begin{equation}
\sigma =0, \ 1,\quad n_{\sigma}=(-1)^{\sigma } \ ,\quad
\nu_{\sigma }=\frac{j+1/2}{\alpha } -\frac{n_{\sigma }}{2} .
\label{nuplusmin}
\end{equation}
On the base of formula (\ref{eigfunc}) we define the positive and
negative frequency eigenfunctions as
\begin{equation}\label{eigfuncpos}
\psi_{\beta } =\left\{ \begin{array}{l} \psi_{\beta }^{(+)}, \quad
{\mathrm{for}} \quad \omega >0 ,\\
\psi_{\beta }^{(-)}, \quad {\mathrm{for}} \quad \omega <0 .
\end{array}
\right.
\end{equation}
These functions are orthonormalized by the condition
\begin{equation}
\int d^{3}x\sqrt{\gamma }\psi_{\beta }^{(\eta
)+}\psi_{\beta^{\prime}}^ {(\eta ^{\prime })}=\delta_{\eta \eta
^{\prime }}\delta_{\beta\beta^ {\prime}} \ , \eta ,\eta ^{\prime
}=\pm ,\label{normcond}
\end{equation}
from which the normalization constant $A$ can be determined.

\section{{Vacuum Expectation Values of the Energy-} \\
Momentum Tensor Inside a Spherical Shell} \label{sec:inside}

In this section we shall consider the vacuum expectation values of
the energy-momentum tensor inside a spherical shell concentric
with the global monopole. The integration in formula
(\ref{normcond}) goes over the interior region of the sphere and
$Z_{\nu}(x)=J_{\nu }(x)$, where $J_{\nu}(x)$ is the Bessel
function of the first kind. We shall assume that on the sphere
surface the field satisfies bag boundary conditions:
\begin{equation}
\left( 1+i\gamma^{\mu }n_{b\mu }\right) \psi =0 \ ,\quad r=a,
\label{boundcond1}
\end{equation}
where $a$ is the sphere radius, $n_{b\mu }$ is the
outward-pointing normal to the boundary. For the sphere $n_{b\mu
}=(0,1,0,0)$. In terms of the spinors $\varphi $ and $\chi $ this
condition is written in the form
\begin{equation}
\varphi +i\left( \vec{\sigma}\cdot\hat{n}_{b}\right)\chi =0 \ ,\quad r=a.
\label{boundcondsp1}
\end{equation}
The imposition of this boundary condition on the eigenfunctions
(\ref {eigfunc}) leads to the following equations for the eigenvalues
\begin{equation}
J_{\nu_{\sigma}+n_{\sigma}}(ka)=n_{\sigma}\frac{\omega
+M}{k}J_{\nu_{\sigma }}(ka) \ . \label{eigenvalues}
\end{equation}
This boundary condition can be written in the form
\begin{equation}\label{boundin}
\tilde{J}_{\nu_{\sigma }}(ka)=0 \ ,
\end{equation}
where now and below for a given function $F(z)$ we shall use the notation
\begin{equation}
\tilde{F}(z)\equiv zF^{\prime }(z)+\left( \mu +s_{\omega
}\sqrt{z^{2}+\mu ^{2}} -(-1)^{\sigma }\nu \right) F(z) \ ,\quad
\sigma =0, \ 1, \label{tildenot}
\end{equation}
with $s_{\omega }={\mathrm{sgn}}(\omega )$ and $\mu=Ma$. Let us denote by
$\lambda _{\nu _{\sigma },s}=ka$, $s=1,2,\ldots ,$ the roots to
equation (\ref{boundin}) in the right half plane, arranged in
ascending order. By taking into account Eq. (\ref{eigenvalues})
and using the standard integral for the Bessel functions, from
condition (\ref{normcond}) for the normalization coefficient one
finds
\begin{equation}
A=A_{\sigma },\quad A_{\sigma }^{-2}\equiv \frac{2\alpha
^{2}a^{2}}{z^{2}}J_{\nu _{\sigma }}^{2}(z)\left[ \left( a\omega
-n_{\sigma } \frac{\nu _{\sigma }}{2}\right) ^{2}-\frac{\nu
_{\sigma }^{2}}{4}-\frac{z^{2}}{2a(\omega +M)}\right] \ ,\quad
z=\lambda _{\nu _{\sigma },s} \ , \label{normcoef}
\end{equation}
with $\omega =\pm \sqrt{\lambda_{l,s}^{(\sigma)2}/a^{2}+M^{2}}$.

Now we expand the field operator in terms of the complete set of
single-particle states $\left\{ \psi_{\beta }^{(+)},\psi_{\beta
}^{(-)}\right\}$:
\begin{equation}\label{operatorexp}
\hat \psi =\sum _{\beta } \left( \hat a_{\beta }\psi_{\beta
}^{(+)} +\hat b_{\beta }^{+}\psi_{\beta }^{(-)}\right) ,
\end{equation}
where $\hat a_{\beta }$ is the annihilation operator for
particles, and $\hat b_{\beta }^{+}$ is the creation operator for
antiparticles. In order to find the vacuum expectation value for
the operator of the energy-momentum tensor we substitute the
expansion (\ref{operatorexp}) and the analog expansion for the
operator $\hat{\bar{\psi }}$ into the corresponding expression for
the spinor fields,
\begin{equation}
T_{\mu \nu }\left\{ \hat{\bar{\psi }},\hat{\psi }
\right\}=\frac{i}{2}\left[ \hat{\bar{\psi}}\gamma_{(\mu }\nabla
_{\nu )}\hat{\psi } -(\nabla_{(\mu }\hat{\bar{\psi}})\gamma_{\nu
)}\hat{\psi } \right] \ . \label{EMTform}
\end{equation}
By making use the standard anticommutation relations for the
annihilation and creation operators, for the vacuum expectation
values one finds the following mode-sum formula
\begin{equation}
\left\langle 0\left| T_{\mu \nu }\right| 0\right\rangle
=\sum_{\beta }T_{\mu\nu }\left\{\bar{\psi }_{\beta
}^{(-)}(x),\psi_{\beta }^{(-)}(x)\right\} \ , \label{modesum}
\end{equation}
where $|0\rangle $ is the amplitude for the corresponding vacuum.
Since the spacetime is spherically symmetric and static, the
vacuum energy-momentum tensor is diagonal, moreover $\langle
T^\theta_\theta\rangle=\langle T^\phi_\phi\rangle$. So in this
case we can write:
\begin{equation}
\left\langle 0\left| T_{\mu }^{\nu }\right| 0\right\rangle ={\rm diag}
(\varepsilon ,-p,-p_{\perp },-p_{\perp }) \ ,
\label{diagform}
\end{equation}
with the energy density $\varepsilon$, radial, $p$, and azimuthal,
$p_{\perp }$, pressures. As a consequence of the continuity
equation $\nabla _\nu \left\langle 0\left| T_{\mu }^{\nu }\right|
0\right\rangle =0$, these functions are related by the equation
\begin{equation}\label{conteq}
r\frac{dp}{dr}+2(p-p_{\perp })=0 \ ,
\end{equation}
which means that the radial dependence of the radial pressure
necessarily leads to the anisotropy in the vacuum stresses.

Substituting eigenfunctions (\ref{eigfunc}) into Eq.
(\ref{modesum}), the summation over the quantum number $m$ can be
done by using standard summation formula for the spherical
harmonics. For the energy-momentum tensor components one finds
\begin{equation}
q(r)=\frac{-1}{8\pi \alpha
^{2}a^{3}r}\sum_{j=1/2}^{\infty}(2j+1)\sum_{\sigma =0,1}
\sum_{s=1}^{\infty }T_{\nu _{\sigma}}(\lambda _{\nu _{\sigma
},s})f_{\sigma\nu_ {\sigma}}^{(q)}\left[ \lambda _{\nu _{\sigma
},s},J_{\nu _{\sigma}} (\lambda _{\nu _{\sigma },s}r/a)\right]
,\quad q=\varepsilon,\ p,\ p_{\perp} \ , \label{qrin}
\end{equation}
where we have introduced the notations
\begin{eqnarray}
f_{\sigma\nu }^{(\varepsilon )}\left[z,J_{\nu
}(y)\right]&=&z\left[ (\sqrt{z^{2}+ \mu^{2}}-\mu )J_{\nu
}^{2}(y)+(\sqrt{z^{2}+\mu ^{2}}+\mu ) J_{\nu +n_{\sigma }}^{2}(y)
\right] \ ,  \label{fnueps} \\
f_{\sigma \nu }^{(p)}\left[ z,J_{\nu }(y)\right]&=&
\frac{z^{3}}{\sqrt{z^{2}+ \mu ^{2}}}\left[ J_{\nu
}^{2}(y)-\frac{2\nu +n_{\sigma }}{y}J_{\nu}(y) J_{\nu +n_{\sigma
}}(y)+J_{\nu +n_{\sigma }}^{2}(y)\right] \ ,
\label{fnup} \\
f_{\sigma \nu }^{(p_{\perp })}\left[ z,J_{\nu }(y)\right] &=&
\frac{z^{3}(2\nu +n_\sigma )}{ 2y\sqrt{z^{2}+\mu ^{2}}}J_{\nu
}(y)J_{\nu +n_{\sigma }}(y) \ . \label{fnupperp}
\end{eqnarray}
Note that in (\ref{qrin}) we have used the relation between the
normalization coefficient and the function $T_{\nu }(z)$
introduced in  Appendix \ref{sec:app1}:
\begin{equation}
A_\sigma^2=\frac{z}{2\alpha^2a^2}\frac{\sqrt{z^2+a^2M^2}+aM}{\sqrt{z^2+a^2M^2}}
T_{\nu_\sigma}(z) \ , \ \ z=\lambda_{\nu _{\sigma },s} \ .
\end{equation}
The vacuum expectation values (\ref{qrin}) are divergent and need
some regularization procedure. To make them finite we can
introduce a cutoff function $\Phi _{\eta }(z)$, $z= \lambda_{\nu
_{\sigma },s}$ with the cutoff parameter $\eta $, which decreases
with increasing $z$ and satisfies the condition $\Phi _{\eta }\to
1$, $\eta \to 0$. Now to extract the boundary-free parts we apply
to the corresponding sums over $s$ the summation formula derived
in Appendix \ref{sec:app1}. As a function $f(z)$ in this formula
we take $f(z)=f_{\sigma\nu_ {\sigma }}^{(q)}\left[z,J_{\nu
_{\sigma }}(zr/a)\right] \Phi _{\eta }(z)$. As a result the
components of the vacuum energy-momentum tensor can be presented
in the form
\begin{equation}
q(r)=q_{m}(r)+q_{b}(r),\quad q=\varepsilon,p,p_{\perp} \ ,
\label{qm+qb}
\end{equation}
where the first term on the right hand side comes from the integral on the
left of summation formula (\ref{sumJ1}), and the second term comes
from the integral on the right of this formula. Making use the
asymptotic formulae for the Bessel modified functions, it can be
seen that for $r<a$ the part $q_b(r)$ is finite in the limit $\eta
\to 0 $ and, hence, in this part the cutoff can be removed. As it
has been pointed out in Appendix \ref{sec:app1}, the function
$f_{\sigma\nu_ {\sigma }}^{(q)}\left[z,J_{\nu _{\sigma
}}(zr/a)\right]$ satisfies relation (\ref{relforf}) and, hence,
the part of the integral on the right of formula (\ref{sumJ1})
over the interval $(0,\mu )$ vanishes after removing the cutoff.
Introducing the notation
\begin{equation}\label{notnunu1}
\nu \equiv \nu _1 =\frac{l}{\alpha } +\frac{1}{2} \ ,
\end{equation}
explicitly summing over $\sigma $ and transforming from summation
over $j$ to summation over $l=j+1/2$, one receives
\begin{equation}\label{qmr}
q_{m}(r)=-\frac{1}{2\pi \alpha ^{2}r}\sum_{l=1}^{\infty} l
\int_{0}^{\infty }\frac{x^3dx}{\sqrt{x^2+M^2}}f_{\nu }^{(q)}\left[
x,J_{\nu }(xr)\right] ,
\end{equation}
where we use the notation
\begin{eqnarray}\label{fnueps1}
f_{\nu }^{(\varepsilon )}\left[ x,J_{\nu }(y)\right] &=& \left( 1+
\frac{M^2}{x^2}\right) \left[ J_{\nu }^2(y) +J_{\nu -1
}^2(y)\right] , \\
f_{\nu }^{(p)}\left[ x,J_{\nu }(y)\right] &=& J_{\nu }^2(y)
+J_{\nu -1}^2(y) -\frac{2\nu }{y}J_{\nu }(y)J_{\nu -1}(y), \label{fnup1}\\
f_{\nu }^{(p_{\perp })}\left[ x,J_{\nu }(y)\right] &=& \frac{\nu
}{y}J_{\nu }(y)J_{\nu -1}(y). \label{fnupperp1}
\end{eqnarray}
Introducing in the expression for $q_{b}(r)$ the modified Bessel
functions, after some transformations end explicitly summing over
$\sigma $, we obtain the formula
\begin{equation}
q_{b}(r)=\frac{1}{\pi ^{2}\alpha ^{2}a^{3}r}\sum_{l=1}^{\infty} l
\int_{\mu }^{\infty}\frac{x^3dx}{\sqrt{x^{2}-\mu^{2}}}
\frac{W\left[ I_{\nu }(x),K_{\nu }(x)\right]}{W\left[ I_{\nu
}(x),I_{\nu }(x)\right]}
 F_{\nu }^{(q)}\left[x,I_{\nu _{\sigma
}}(xr/a)\right] \ , \label{qb}
\end{equation}
with
\begin{eqnarray}\label{Fnueps}
F_{\nu }^{(\varepsilon )}\left[x,I_{\nu }(y)\right] &=& \left( 1-
\frac{\mu ^2}{x^2}\right) \left\{ I_{\nu -1}^2(y)-I_{\nu
}^2(y)-\mu \frac{I_{\nu -1}^2(y)+I_{\nu
}^2(y)}{W\left[ I_{\nu }(x),K_{\nu }(x)\right]}\right\} ,\\
F_{\nu }^{(p)}\left[x,I_{\nu }(y)\right] &=& I_{\nu
-1}^2(y)-I_{\nu }^2(y)-\frac{2\nu -1}{y}I_{\nu }(y)I_{\nu -1}(y),
\label{Fnup} \\
F_{\nu }^{(p_{\perp })}\left[x,I_{\nu }(y)\right] &=& \frac{\nu
-1/2}{y}I_{\nu }(y)I_{\nu -1}(y) \ . \label{Fnupperp}
\end{eqnarray}
Here and below for given functions $f(x)$ and $g(x)$ we use the
notation
\begin{equation}\label{Wnot}
W\left[ f(x),g(x)\right] =\left[ xf'(x)+(\mu +\nu )f(x)\right]
\left[ xg'(x)+(\mu +\nu )g(x)\right] +(x^2-\mu ^2)f(x)g(x).
\end{equation}
As we see, the part $q_{m}(r)$ in the vacuum expectation value for
the energy-momentum tensor do not depend on the radius of the
sphere $a$, whereas the contribution of the terms $q_{b}(r)$ tends
to zero as $a\rightarrow\infty $ (for large $a$ the subintegrand
behaves as $e^{2x(r/a-1)}$). It follows from here that  the
quantities (\ref{qmr}) are the vacuum expectation values for the
components of the energy-momentum tensor for the unbounded global
monopole space:
\begin{equation}
\langle 0_{m}|T_{\mu}^{\nu }|0_{m}\rangle ={\rm diag}(\varepsilon_{m},-p_{m},-p_{m},
-p_{\perp m}) \ ,
\label{EMTunbound}
\end{equation}
where $|0_{m}\rangle $ is the amplitude for the corresponding
vacuum. Note that in expressions (\ref{qmr}) for the corresponding
components we have not explicitly written the cutoff function.
Precisely speaking in this form all terms related with (\ref{qmr})
are divergent. The renormalization prescriptions adopted to
provide a finite and well defined result to them are the usual
ones applied for the curved spacetime without boundary \cite{Chr,
Wald,Birr82}. For the specific system analyzed here the
point-splitting renormalization procedure has been applied in
previous publication \cite{EVN}. The part $q_{b}(r)$ in Eq.
(\ref{qm+qb}) is induced by the presence of the spherical shell
and can be termed as the boundary part. As we have seen, the
application of the generalized Abel-Plana formula allows us to
extract from the vacuum expectation value of the energy-momentum
tensor the contribution due to the boundary-free monopole
spacetime and to present the boundary-induced part in terms of
exponentially convergent integrals (for applications of the
generalized Abel-Plana formula to a number of Casimir problems
with various boundary geometries see Refs.
\cite{Saharian,Saha00,Rome02,Saha01,Rome01,Saha02,Avag02}). It can
be easily checked that the both terms on the right of formula
(\ref{qm+qb}), $q_{m}(r)$ and $q_{b}(r)$, obey the continuity
equation (\ref{conteq}). In addition, as it is seen from
expressions (\ref{qb})--(\ref{Fnupperp}), for a massless spinor
field the boundary-induced part of the vacuum energy-momentum
tensor is traceless and the trace anomalies are contained only in
the purely global monopole part without boundaries.

Having the components of the energy-momentum tensor we can find
the corresponding fermionic condensate $\langle 0 |\bar \psi \psi
|0\rangle $ making use the formula for the trace of the
energy-momentum tensor, $T_\mu ^\mu =M\bar \psi \psi $. It is
presented in the form of the sum
\begin{equation}\label{condens}
\langle 0 |\bar \psi \psi |0\rangle =\langle 0_m |\bar \psi \psi
|0_m \rangle +\langle \bar \psi \psi \rangle _b \ ,
\end{equation}
where the boundary-free part (first summand on the right) and the
sphere-induced part (second summand on the right) are determined
by formulae similar to Eqs. (\ref{qmr}) and (\ref{qb}),
respectively, with replacements
\begin{eqnarray}\label{fnucondens}
f_{\nu }^{(q)}\left[ x,J_{\nu }(y)\right] &\to &
\frac{M}{x^2}\left[ J_{\nu }^2(y)+J_{\nu -1}^2(y)\right] ,\\
F_{\nu }^{(q)}\left[ x,I_{\nu }(y)\right] &\to &
-\frac{a}{x^2}\left\{ \mu \left[ I_{\nu -1}^2(y)-I_{\nu
}^2(y)\right] +(x^2-\mu ^2) \frac{I_{\nu -1}^2(y)+I_{\nu
}^2(y)}{W\left[ I_\nu (x),K_\nu (x)\right]}\right\} \ ,
\label{Fnucondens}
\end{eqnarray}
with notation (\ref{Wnot}). Alternatively one could obtain
formulae (\ref{condens})--(\ref{Fnucondens}) applying the
summation formula (\ref{sumJ1}) to the corresponding mode-sum
$\sum_{\beta} \bar \psi _{\beta }^{(-)} \psi _{\beta }^{(-)}$ for
the fermionic condensate.

Note that formulae (\ref{qm+qb}), (\ref{qmr}), (\ref{qb}) can be
obtained by another equivalent way, applying a certain first-order
differential operator on the corresponding Green function and
taking the coincidence limit. To construct the Green function we
can use the corresponding mode expansion formula with the
eigenfunctions (\ref{eigfunc}). This function is a $4\times 4$
matrix and the angular parts of the corresponding elements are
products of the components for the spinor spherical harmonics,
$\Omega _{jlm}^{(n)}(\theta ,\phi ) \Omega _{jlm}^{(n')+}(\theta '
,\phi ')$, where the upper indices $n,n'=1,2$ numerate the spinor
components. These parts are the same as in the boundary-free case
and coincide with the corresponding functions for the Minkowski
bulk. The radial parts for the components of the Green function
contain the products of the Bessel functions in the forms $J_{\nu
_{\sigma }+\tau n_{\sigma }}(zr/a) J_{\nu _{\sigma }+\tau '
n_{\sigma }}(zr'/a)$, $\tau , \tau '=0,1$, where $z=\lambda _{\nu
_{\sigma },s}$. To evaluate the sum over $s$ we can apply the
summation formula (\ref{sumJ1}). The condition (\ref{condf}) is
satisfied if $r+r'+|t-t'|<2a$. The term with the integral on the
left of formula (\ref{sumJ1}) gives the Green function for the
boundary-free global monopole spacetime, and the term with the
integral on the right will give the boundary-induced part.

\begin{figure}[tbph]
\begin{center}
\epsfig{figure=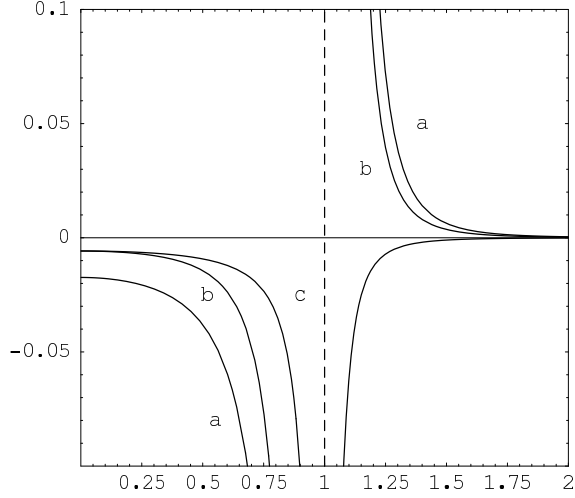,width=7.5cm,height=6.5cm}
\end{center}
\caption{Vacuum energy density, $a^4\varepsilon $ (curve a),
azimuthal pressure $a^4p_{\perp }$ (curve b), and radial pressure
$a^4p$ (curve c) for a massless spinor as functions on the ratio
$r/a$ inside and outside a spherical shell in the Minkowski
spacetime ($\alpha =1$).} \label{fig1Mink}
\end{figure}

In the case $\alpha =1$ the quantities (\ref{qmr}) present the
vacuum expectation values for the Minkowski spacetime without
boundaries. This can be also seen by direct evaluation. For
example, in the case of the energy density making use the formula
$\sum_{l=0}^{\infty }(2l+1)J_{l+1/2}^2(y)=2y/\pi $, one finds
\begin{eqnarray}\label{Minken}
\varepsilon _m(r)&=&-\frac{1}{2\pi r}\sum_{l=0}^{\infty
}l\int_{0}^{\infty } dx\, x^2\sqrt{x^2+M^2}\left[
J_{l+1/2}^2(xr)+J_{l-1/2}^2(xr)\right] \nonumber \\
&=& -2\int \frac{d^3k}{(2\pi )^3}\sqrt{k^2+M^2},
\end{eqnarray}
which is precisely the energy density of the Minkowski vacuum for
a spinor field. As for the Monkowski background the renormalized
vacuum energy-momentum tensor vanishes,
$q_{m}(r)_{{\mathrm{ren}}}=0$, the vacuum energy-momentum tensor
is purely boundary-induced and the corresponding components are
given by formulae (\ref{qb})--(\ref{Fnupperp}) with $\nu =l+1/2$.
Note that the previous investigations on the spinor Casimir effect
for a spherical boundary (see, for instance,
\cite{Bend76,Milt80,Milt83,Baac83,Blau88,Eliz98,Cogn01,Most97,Plun86,Milt02,Bord01}
and references therein) consider mainly global quantities, such as
total vacuum energy. For the case of a massless spinor  the
density of the fermionic condensate $\langle \bar \psi \psi
\rangle _b$ is investigated in \cite{Milt81} (see also
\cite{Milt02}). The corresponding formula derived in Ref.
\cite{Milt81} is obtained from (\ref{qb}) with replacement
(\ref{Fnucondens}) in the limit $\mu = 0$. In Fig. \ref{fig1Mink}
we have presented the dependence of the Casimir densities,
$a^4q_b(r)$ on the rescaled radial coordinate $r/a$ for a massless
spinor field on the Minkowski bulk. The vacuum energy density and
pressures are negative inside the sphere.

Now we turn to the consideration of various limiting cases of the
expressions for the sphere-induced vacuum expectation values. In
the limit $r\to 0$, for the boundary parts (\ref{qb}) the summands
with a given $l$ behave as $r^{2l/\alpha -2}$, and the leading
contributions come from the lowest $l=1$ terms. Making use
standard formulae for the Bessel modified functions for small
values of the argument, for the sphere-induced parts near the
center, $r\ll a$, one finds
\begin{eqnarray}\label{epsbrto0}
\varepsilon _b &\approx &\frac{\pi ^{-2}a^{-4}}{2\alpha ^2\Gamma
^2\left( \frac{1}{\alpha }+\frac{1}{2}\right)}\left(
\frac{r}{2a}\right) ^{\frac{2}{\alpha }-2}\int_{\mu }^{\infty }dx
\, x^{\frac{2}{\alpha }}\sqrt{x^2-\mu ^2}\frac{W\left[ I_{\nu }
(x),K_{\nu }(x)\right] -\mu }{W\left[ I_{\nu }(x),I_{\nu
}(x)\right] } \ , \\
p_b &\approx & \alpha p_{\perp b}\approx \frac{\pi
^{-2}a^{-4}}{2\alpha (2+\alpha )\Gamma ^2\left( \frac{1}{\alpha
}+\frac{1}{2}\right)}\left( \frac{r}{2a}\right) ^{\frac{2}{\alpha
}-2}\int_{\mu }^{\infty }\frac{x^{\frac{2}{\alpha
}+2}dx}{\sqrt{x^2-\mu ^2}} \frac{W\left[ I_{\nu }(x),K_{\nu
}(x)\right] }{W\left[ I_{\nu }(x),I_{\nu }(x)\right] } ,
\label{pbrto0}
\end{eqnarray}
where $\nu =1/\alpha +1/2$ and $\Gamma (x)$ is the gamma function.
Hence, at the sphere center the boundary parts vanish for the
global monopole spacetime ($\alpha <1$) and are finite for the
Minkowski spacetime ($\alpha =1$). Note that in the large mass
limit, $\mu \gg 1$, the integrals in Eqs. (\ref{epsbrto0}),
(\ref{pbrto0}) are exponentially suppressed by the factor
$e^{-2\mu }$. In the Minkowski background case the vacuum stresses
for a massless spinor are isotropic at the sphere center and after
the numerical evaluation of the integral one finds
\begin{equation}\label{pb(0)}
p_b(0)=p_{\perp b}(0)=\frac{\varepsilon _b }{3}=\frac{2}{3\pi
^2a^4}\int_{0}^{\infty }\frac{e^{-2x}(x^2+x-e^x\sinh x)dx
}{(2x^2+1)\cosh (2x)-2x\sinh (2x)-1} = -\frac{0.00579}{a^4} \ .
\end{equation}

\begin{figure}[tbph]
\begin{center}
\epsfig{figure=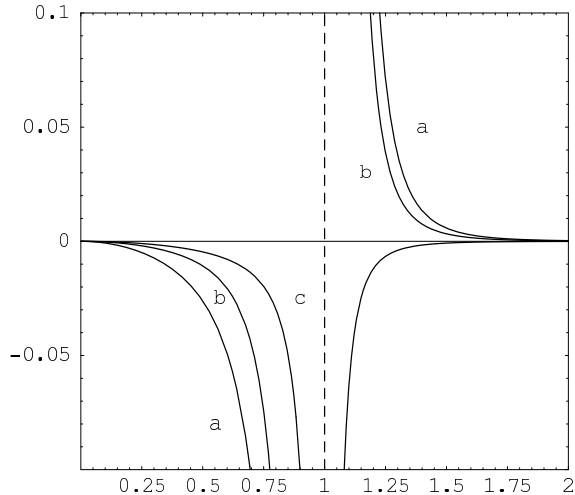,width=7.5cm,height=6.5cm}
\end{center}
\caption{The same as in Fig. \ref{fig1Mink} for the boundary-induced parts
$a^4q_b(r)$ on background of the global monopole spacetime with
$\alpha =0.5$.} \label{fig2mon}
\end{figure}

The boundary induced parts of the vacuum energy-momentum tensor
components diverge at the sphere surface, $r\to a$. These
divergences are well-known in quantum field theory with boundaries
and are investigated for various types of boundary geometries
\cite{Bali78,Deut79,Kenn80}. In order to find the leading terms of
the corresponding asymptotic expansion in powers of the distance
from the sphere surface, we note that in the limit $r\to a$ the
sum over $l$ in (\ref{qb}) diverges and, hence, for small $1-r/a$
the main contribution comes from the large values of $l$.
Consequently, rescaling the integration variable $x\to \nu x$ and
making use the uniform asymptotic expansions for the modified
Bessel functions for large values of the order \cite{Abra64}, to
the leading order one finds
\begin{eqnarray}\label{epsbas}
\varepsilon _b(r) &\approx & -\frac{\mu +1/5}{12 \pi ^2a(a-r)^3},\\
p_b(r) &\approx & \left( 1-\frac{r}{a}\right) p_{\perp
b}(r)\approx -\frac{1/5-2\mu }{24 \pi ^2a^2(a-r)^2} .\label{pbas}
\end{eqnarray}
Notice that the terms in these expansions diverging as the inverse
fourth power of the distance have cancelled out. This is a
consequence of the conformal invariance of the massless fermionic
field and is in agreement with general conclusions of Ref.
\cite{Deut79}. Near the sphere surface the energy density is
negative for all values of $\mu$, while the vacuum pressures are
negative for $\mu <0.1$ and are positive for $\mu >0.1$. It is of
interest to note that the leading terms do not depend on the
parameter $\alpha $ and, hence, are the same for the Minkowski and
global monopole bulks. For the latter case due to the divergences,
near the sphere surface the total vacuum energy-momentum tensor is
dominated by the boundary induced parts $q_b(r)$. The dependence
of these parts on the rescaled radial coordinate $r/a$ is depicted
in Fig. \ref{fig2mon} for the case of a massless fermionic field
on the global monopole background with the solid angle deficit
parameter $\alpha =0.5$.

\begin{figure}[tbph]
\begin{center}
\begin{tabular}{cc}
\epsfig{figure=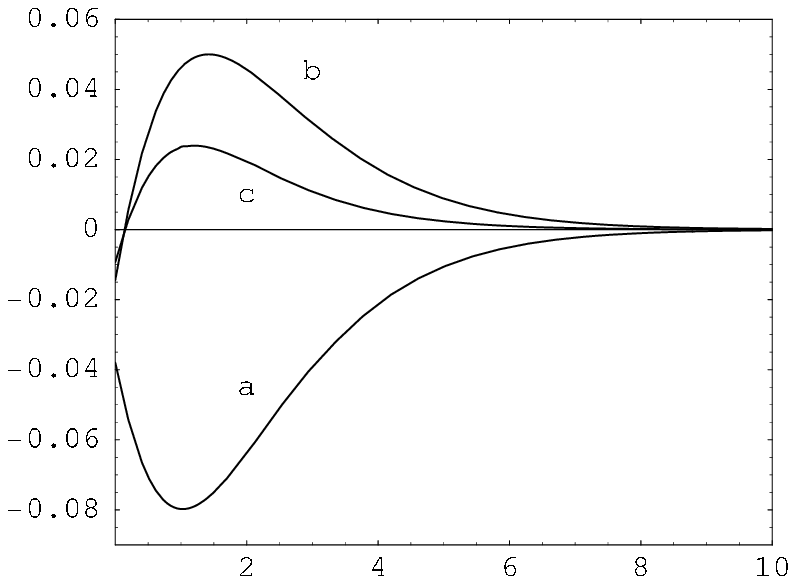,width=6.5cm,height=6cm}& \quad
\epsfig{figure=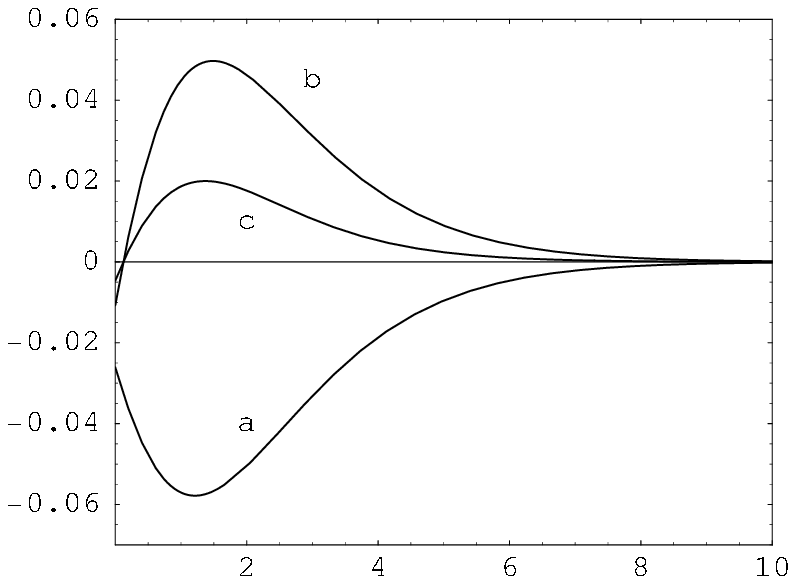,width=6.5cm,height=6cm}
\end{tabular}
\end{center}
\caption{Boundary-induced vacuum expectation values $a^4q_b(r)$,
$q=\varepsilon ,p,p_{\perp }$, as functions on $\mu =Ma$ for
$r/a=0.5$. The curves a, b, c correspond to the energy density
($\varepsilon $), azimuthal pressure ($p_{\perp }$), and radial
pressure ($p$), respectively. For the left panel $\alpha =1$
(Monkowski spacetime) and for the right panel $\alpha =0.5$.}
\label{fig3mass}
\end{figure}

Now let us consider the limit $\alpha \ll 1$ for a fixed value
$r<a$. This limit corresponds to strong gravitational fields. In
this case from (\ref{notnunu1}) one has $\nu \approx l/\alpha \gg
1$, and after introducing in (\ref{qb}) a new integration variable
$y=x/\nu $, we can replace the modified Bessel functions by their
uniform asymptotic expansions for large values of the order. The
integral over $y$ can be estimated by making use the Laplace
method. The main contribution to the sum over $l$ comes from the
summands with $l=1$ and the boundary parts of the vacuum
energy-momentum tensor components behave as $\exp [-2\ln
(a/r)/\alpha ]$ with $p_b/p_{\perp b}\sim \alpha $. Hence, for
$\alpha \ll 1$ the boundary-induced vacuum expectation values are
exponentially suppressed and the corresponding vacuum stresses are
strongly anisotropic. Fig. \ref{fig3mass} shows that the nonzero
mass can essentially change the behavior of the vacuum
energy-momentum tensor components. In this figure we have depicted
the dependence of the boundary induced quantities $a^4q_b(r)$ on
the parameter $Ma$ for the radial coordinate $r=0.5a$. The left
panel corresponds to the sphere in the Minkowski spacetime
($\alpha =1$) and for the right panel $\alpha =0.5$.

\section{Vacuum Expectation Values Outside a Spherical Shell} \label{sec:outside}

Now let us consider the expectation values of the energy--momentum
tensor in the region outside a spherical shell, $r>a$. The
corresponding eigenfunctions have the form (\ref{eigfunc}), where
now the function $Z_{\nu }(kr)$ is the linear combination of the
Bessel functions of the first and second kinds. The coefficient in
this linear combination is determined from the boundary condition
(\ref{boundcondsp1}) and one obtains
\begin{equation}\label{gnu}
Z_{\nu }(kr)=g_{\nu }(ka,kr)\equiv J_{\nu }(kr)\tilde{Y}_{\nu
}(ka)- Y_{\nu }(kr)\tilde{J}_{\nu }(ka) \ ,
\end{equation}
where $Y_{\nu }(z)$ is the Bessel function of the second kind, and
the functions with tilda are defined as (\ref{tildenot}). Now the
spectrum for the quantum number $k$ is continuous and the
corresponding $\delta _{kk'}$ in Eq. (\ref{normcond}) is
understood as the Dirac delta function $\delta (k-k')$. To find
the normalization coefficient $A$ from Eq. (\ref{normcond}) it is
convenient to take $\beta =\beta '$ for all discrete quantum
numbers. As the normalization integral diverges in the limit
$k=k'$, the main contribution into the integral over radial
coordinate comes from large values of $r$ when the Bessel
functions can be replaced by their asymptotics for large
arguments. The resulting integral is taken elementary and for the
normalization coefficient we obtain
\begin{equation}\label{Aout}
A=A_{\sigma },\quad A_{\sigma }^2=\frac{k(\omega
+M)}{2\alpha^2\omega [\tilde{J}_ {\nu_{\sigma
}}^2(ka)+\tilde{Y}_{\nu_{\sigma }}^2(ka)]} \ .
\end{equation}
Substituting the eigenfunctions (\ref{eigfunc}) into the mode-sum
formula (\ref{modesum}) and taking into account Eqs. (\ref{gnu})
and (\ref{Aout}), we can see that the vacuum energy-momentum
tensor has the form (\ref{diagform}). The diagonal components are
determined by formulae
\begin{equation}\label{qrout}
q(r)=\frac{-1}{8\pi \alpha
^2a^3r}\sum_{j=1/2}^{\infty}(2j+1)\sum_{\sigma =0,1}
\int_{0}^{\infty }dx\frac{f_{\sigma \nu_\sigma
}^{(q)}\left[x,g_{\nu _{\sigma}}(x,xr/a)\right]}{\tilde{J}_{\nu
_{\sigma}}^2(x)+\tilde{Y}_ {\nu _{\sigma }}^2(x)} \ ,\quad
q=\varepsilon, \ p, \ p_{\perp } \ ,
\end{equation}
where the expressions for $f_{\sigma \nu _\sigma }^{(q)}\left[x,g_{\nu _{\sigma }}
(x,xr/a)\right]$ are obtained from formulae
(\ref{fnueps})--(\ref{fnupperp}) by replacements
\begin{equation}\label{Jtog}
J_{\nu }(y)\to g_{\nu }(x,y)\ ,\quad J_{\nu +n_\sigma }(y)\to
J_{\nu+n_\sigma }(y) \tilde{Y}_{\nu }(ka)-Y_{\nu
+n_\sigma}(y)\tilde{J}_{\nu }(ka) \ .
\end{equation}
To find the parts in the vacuum expectation values of the
energy-momentum tensor induced by the presence of the sphere we
subtract the corresponding components for the monopole bulk
without boundaries, given by Eq. (\ref{qmr}). In order to evaluate
the corresponding difference we use the relation
\begin{equation}
\label{relf} \frac{f_{\sigma \nu _\sigma }^{(q)}\left[ x,g_{\nu
_{\sigma}}(x,xr/a)\right]} {\tilde{J}_{\nu
_{\sigma}}^2(x)+\tilde{Y}_{\nu _{\sigma }}^2(x)}-f_ {\sigma
\nu_\sigma}^{(q)}\left[ x,J_{\nu _{\sigma }}(xr/a)\right]
=-\frac{1}{2}\sum_{s=1,2}\frac{\tilde{J}_{\nu
_{\sigma}}(x)}{\tilde{H}_ {\nu _{\sigma }}^{(s)}(x)}f_{\sigma
\nu_\sigma }^{(q)}\left[ x,H_{\nu_{\sigma }}^ {(s)}(xr/a)\right] \
,
\end{equation}
where $H_{\nu }^{(s)}(z)$, $s=1, \ 2$ are the Hankel functions.
This allows to present the vacuum energy-momentum tensor
components in the form (\ref{qm+qb}) with the boundary induced
parts
\begin{equation}\label{qrbout}
q_b(r)=\frac{1}{16\pi \alpha
^2a^3r}\sum_{j=1/2}^{\infty}(2j+1)\sum_{\sigma =0,1}
\sum_{s=1,2}\int_{0}^{\infty}dx\frac{\tilde{J}_{\nu _{\sigma
}}(x)} {\tilde{H}_{\nu_{\sigma }}^{(s)}(x)}f_{\sigma \nu _\sigma
}^{(q)} \left[x,H_{\nu_{\sigma }}^{(s)}(xr/a)\right] \ .
\end{equation}
On the complex plane $x$ we can rotate the integration contour on
the right of this formula by the angle $\pi /2$ for $s=1$ and by
the angle $-\pi /2$ for $s=2$. The integrals over the segments
$(0,i\mu )$ and $(0,-i\mu )$ cancel out and after introducing the
Bessel modified functions one obtains
\begin{equation}
q_{b}(r)=\frac{1}{\pi ^{2}\alpha ^{2}a^{3}r}\sum_{l=1}^{\infty} l
\int_{\mu }^{\infty}\frac{x^3dx}{\sqrt{x^{2}-\mu^{2}}}
\frac{W\left[ I_{\nu }(x),K_{\nu }(x)\right]}{W\left[ K_{\nu
}(x),K_{\nu }(x)\right]}
 F_{\nu }^{(q)}\left[x,K_{\nu }(xr/a)\right] . \label{qbout}
\end{equation}
Here the expressions for the functions $F_{\sigma\nu}^{(q)}
\left[x,K_{\nu }(y)\right]$ are obtained from formulae
(\ref{Fnueps})--(\ref{Fnupperp}) by replacements $I_\nu (y)\to
K_{\nu }(y)$ and $I_{\nu -1}(y)\to -K_{\nu -1}(y)$. With the same
replacements, from (\ref{condens}), (\ref{fnucondens}),
(\ref{Fnucondens}) we can obtain formulae for the fermionic
condensate in the region outside a sphere. As for the interior
region, it can be seen that in the limit of strong gravitational
fields, $\alpha \ll 1$, the boundary induced vacuum expectation
values are exponentially suppressed by the factor $\exp
[-(2/\alpha )\ln (r/a)]$, an the corresponding vacuum stresses are
strongly anisotropic: $p_b/p_{\perp b}\sim \alpha $.

In the case $\alpha =1$ the renormalized values for the boundary-%
free parts $q_m(r)$ vanish and from (\ref{qbout}) with $\nu
=l+1/2$ the components of the vacuum energy-momentum tensor are
obtained for the region outside a spherical shell on the Minkowski
bulk. Previous approaches to this problem have been global and our
calculation is a local extension of these results. Note that for
an infinitely thin spherical shell the total vacuum energy for a
massless spinor, including interior and exterior parts, is
positive, $E=0.0204/a$ \cite{Milt83,Eliz98,Cogn01}. In Fig.
\ref{fig1Mink} we have plotted the dependence of the vacuum energy
density and stresses on the radial coordinate for a massless
spinor field outside a sphere on the Minkowski bulk. The same
graphics for the boundary-induced expectation values (\ref{qbout})
on background of the global monopole spacetime with $\alpha =0.5$
are depicted in Fig. \ref{fig2mon}. As seen form these figures,
the energy density and azimuthal pressure are positive outside a
sphere, and the radial pressure is negative. The latter has the
same sign as for the interior region.

For the case of a massless spinor the asymptotic behavior of
boundary part (\ref{qbout}) at large distances from the sphere can
be obtained by introducing a new integration variable $y=xr/a$ and
expanding the subintegrands in terms of $a/r$. The leading
contribution for the summands with a given $l$ has an order
$(a/r)^{2\nu +4}$ and the main contribution comes from the $l=1$
term. Evaluating the standard integrals involving the square of
the Mac-Donald function, the leading terms for the asymptotic
expansions over $a/r$ can be presented in the form
\begin{equation}\label{qblarger}
q_b(r)\approx \frac{1}{2^{\frac{2}{\alpha }}\pi a^4}\frac{\Gamma
\left(\frac{1}{\alpha }+1\right) \Gamma\left( \frac{2}{\alpha }
+\frac{3}{2}\right) f_q}{(4-\alpha ^2)(2+\alpha )\Gamma ^3\left(
\frac{1}{\alpha } +\frac{1}{2}\right) } \left( \frac{a}{r}\right)
^{\frac{2}{\alpha }+5} \ ,
\end{equation}
where
\begin{equation}\label{fqlarge}
f_{\varepsilon }=4\frac{\alpha +1}{3\alpha +2},\quad
f_p=-\frac{2\alpha }{3\alpha +2}, \quad f_{p_{\perp }}=1.
\end{equation}

As for the interior components, the quantities (\ref{qbout})
diverge at the sphere surface $r=a$. Near the surface the dominant
contributions come from modes with large $l$ and by making use the
uniform asymptotic expansions for the Bessel modified functions,
the asymptotic expansions can be derived in powers of the distance
from the sphere. The leading terms of these asymptotic expansions
are determined by formulae
\begin{eqnarray}\label{epsbasout}
\varepsilon _b(r) &\sim & \frac{ 1/5-5\mu }{12 \pi ^2a(r-a)^3},\\
p_b(r) &\sim & -\left( \frac{r}{a}-1\right) p_{\perp b}(r)\sim
-\frac{1/5-2\mu }{24 \pi ^2a^2(r-a)^2} .\label{pbasout}
\end{eqnarray}
Near the sphere the outside energy density is positive for $\mu
<0.04$ and is negative for $\mu >0.04$. Recall that near the
sphere the interior energy density is always negative. As we see,
the leading terms for the radial pressure are the same for the
regions outside and inside the sphere. For the azimuthal pressure
these terms have opposite signs. In the case of the massless
spinor field the same is true for the energy density.

\section{Concluding Remarks} \label{sec:conc}

In this paper we have analyzed the fermionic Casimir densities induced by a
spherical shell in a idealized point-like global monopole spacetime. Specifically
the renormalized vacuum expectation value of the operator energy-momentum tensor
has been considered when the matter fields obey the MIT bag boundary condition on
the shell. Because the boundary condition provides a discrete energy spectrum for the
matter fields, the summation over the modes can be developed by using the the
generalized Abel-Plana summation formula. Moreover, this procedure allows us to
extract from the vacuum average the boundary dependent part. This part
presents, besides the contribution coming from the parameter $\alpha$ which
characterizes the presence of the global monopole, contributions coming from the
boundary itself. Two distinct situations have been considered: the calculation of
the vacuum average in the regions inside and outside of the spherical shell. As to
the inside region case, we pointed out that all contributions to the vacuum average
go to zero as the radius of the spherical shell goes to infinity, as it was expected.

The boundary-induced expectation values for the components of the
energy-momentum tensor are given by formulae (\ref{qb}) and
(\ref{qbout}) for interior and exterior regions, respectively. The
corresponding formulae for the fermionic condensate densities are
obtained from these expressions with replacement
(\ref{Fnucondens}) for the interior region and with additional
replacement $I_{\nu }(y)\to K_{\nu }(y)$, $I_{\nu -1}(y)\to
-K_{\nu -1}(y)$ for the exterior region. These expressions diverge
in a non-integrable manner as the boundary is approached. The
energy density and azimuthal pressure vary, to leading order, as
the inverse cube of the distance from the sphere, and near the
sphere the azimuthal pressure has opposite signs for the interior
and exterior regions. For a massless spinor the same is true for
the energy density. The radial pressure varies as the inverse
square of the distance and near the sphere has the same sign for
exterior and interior regions. This behavior is clearly seen from
Figs. \ref{fig1Mink} and \ref{fig2mon} where the radial dependence
of the vacuum energy density, azimuthal and radial pressures are
presented for the Minkowski background and global monopole
spacetime with $\alpha =0.5$. The leading terms of the
corresponding asymptotic expansions near the sphere do not depend
on the solid angle deficit parameter and are the same for these
two cases. Near the sphere the interior energy density is negative
for all values of the mass, while the exterior energy density is
positive for $Ma<0.04$ and is negative for $Ma>0.04$. in order to
illustrate the dependence on the mass, in Fig. \ref{fig3mass} we
have plotted the boundary-induced vacuum densities at $r=0.5a$ as
functions on $Ma$. Near the sphere center the dominant
contributions come from modes with $l=1$ and the sphere-induced
vacuum expectation values vanish for the global monopole spacetime
and are finite for the Minkowski bulk. The asymptotic behavior of
the vacuum energy-momentum tensor at large distances from the
sphere is described by formula (\ref{qblarger}) and the
corresponding diagonal components go to zero as $(a/r)^{2/\alpha
+5}$. In the limit of strong gravitational field, corresponding to
small values of the parameter $\alpha $, describing the solid
angle deficit, the boundary-induced part of the vacuum
energy-momentum tensor is strongly suppressed by the factor $\exp
[-(2/\alpha ) |\ln (r/a)|]$ and the corresponding vacuum stresses
are strongly anisotropic: $p_b\sim \alpha p_{\perp b}$. Note that
this suppression effect also takes place in the scalar case
\cite{A-M,Saha03b}.

\section*{Acknowledgement}
AAS acknowledges the hospitality of the Abdus Salam International
Centre for Theoretical Physics, Trieste, Italy. He was supported
in part by the Armenian Ministry of Education and Science Grant
No. 0887. ERBM wants to thank Conselho Nacional de Desenvolvimento
Cient\'\i fico e Tecnol\'ogico (CNPq.) for partial financial
support.

\appendix

\section{Summation Formula over the Zeros of a Combination of the Bessel
Functions} \label{sec:app1}

We have seen that the vacuum expectation values for the
energy-momentum tensor for a spinor field inside a spherical shell
on background of the global monopole spacetime contain sums over
the zeros of the function
\begin{equation}
\tilde{J}_{\nu }(z)\equiv zJ_{\nu }^{\prime }(z)+\left( \mu
+s_{\omega }\sqrt{ z^{2}+\mu ^{2}}-(-1)^{\sigma }\nu \right)
J_{\nu }(z) \ ,\quad \sigma=0,\ 1 \ ,\ s_{\omega }=\pm 1 \ .
\label{Jtilde1}
\end{equation}
To obtain a summation formula over these zeros, we use here the
generalized Abel-Plana formula \cite{Saharian,Saha00}. In this
formula, as a function $g(z)$ let us choose
\begin{equation}
g(z)=i\frac{\tilde{Y}_{\nu }(z)}{\tilde{J}_{\nu }(z)}f(z) \ ,
\label{gebessel}
\end{equation}
with a function $f(z)$ analytic in the right half-plane
${\mathrm{Re}}z\geq 0$, and $Y_{\nu }(z)$ is the Neumann function. For the sum and
difference of the functions $f(z)$ and $g(z)$ one obtains
\begin{equation}
f(z)-(-1)^{k}g(z)=\frac{\tilde{H}_{\nu}^{(k)}(z)}{\tilde{J}_{\nu
}(z)}f(z) \ , \quad k=1, \ 2 \ , \label{gefsum}
\end{equation}
with $H_{\nu }^{(1)}$ and $H_{\nu }^{(2)}$ being Bessel functions
of the third kind or Hankel functions. By using the asymptotic
formulae for the Bessel functions for large values of the
argument, the conditions for the generalized Abel-Plana formula
can be written in terms of the function $f(z)$ as follows:
\begin{equation}
|f(z)|<\epsilon (x)e^{c|y|} \ ,\quad z=x+iy,\quad |z|\rightarrow
\infty  \ ,
\label{condf}
\end{equation}
where $c<2$  and $\epsilon (x)\rightarrow 0$ for $x\rightarrow\infty $.

Let  $\lambda _{\nu ,s}\neq 0$, $s=1,2,3\ldots $ be zeros for the
function $\tilde{J}_{\nu }(z)$ in the right half-plane, arranged
in ascending order, $\lambda _{\nu ,s}\leq \lambda _{\nu ,s+1}$.
By using the Wronskian $W[J_{\nu }(z),Y_{\nu }(z)]=2/\pi z$, one
can easily see that $\tilde{Y}(\lambda _{\nu ,s})=2/\left( \pi
\tilde{J} (\lambda _{\nu ,s})\right) $. This allows to present the
residue term coming from the poles of the function $g(z)$ in the
form
\begin{equation}
\pi i{\mathrm{Res}}_{z=\lambda _{\nu ,s}}g(z)=T_{\nu }(\lambda
_{\nu ,s})f(\lambda _{\nu ,s}) \ ,
\label{rbessel}
\end{equation}
where we have introduced the notation
\begin{equation}
T_{\nu }(z)=\frac{z}{J_{\nu }^{2}(z)\left[ z^{2}+(\mu-(-1)^{\sigma
}\nu ) (\mu ^{2}+s_{\omega }\sqrt{z^{2}+\mu^{2}})-\frac{s_{\omega
}z^{2}}{2\sqrt{z^{2}+\mu ^{2}}}\right] } \ . \label{r1}
\end{equation}
Substituting (\ref{gebessel}) and (\ref{gefsum}) into the
generalized Abel-Plana formula \cite{Saharian,Saha00} and taking
in this formula the limit $a\rightarrow 0$ (the branch points
$z=\pm i\mu $ are avoided by semicircles of small radius), we
obtain that for the function $f(z)$ analytic in the half-plane $
{\rm Re}z>0$ and satisfying condition (\ref{condf}) the following
formula takes place
\begin{eqnarray}
&&{}\lim_{b\rightarrow +\infty }\left\{ \sum_{s=1}^{n}T_{\nu}(\lambda _{\nu,s})
f(\lambda _{\nu ,s})-\int_{0}^{b}f(x)dx\right\} =\frac{\pi}{2}
{\mathrm{Res}}_{z=0}f(z)\frac{\bar{Y}_{\nu}(z)}{\bar{J}_{\nu}(z)}  \nonumber \\
&&{}-\frac{1}{\pi }\int_{0}^{\infty }\left[ e^{-\nu \pi i}f(xe^{\pi i/2})
\frac{K_{\nu}^{(+)}(x)}{I_{\nu }^{(+)}(x)}+e^{\nu \pi i}f(xe^{-\pi i/2})
\frac{K_{\nu}^{(-)}(x)}{I_{\nu }^{(-)}(x)}\right] dx \ ,
\label{sumJ1}
\end{eqnarray}
where on the left $\lambda _{\nu,n}<b<\lambda_{\nu,n+1}$,  and
$T_{\nu }(\lambda _{\nu ,k})$ is determined by relation
(\ref{r1}). In formula (\ref{sumJ1})\bigskip\ we use the notation
\begin{equation}
F^{(\pm )}(z)=\left\{
\begin{array}{cc}
zF^{\prime }(z)+\left( \mu +s_{\omega }\sqrt{\mu
^{2}-z^{2}}-(-1)^{\sigma}\nu \right)
F(z) \ ,\quad  & |z|<\mu  \\
zF^{\prime }(z)+\left(\mu\pm s_{\omega
}i\sqrt{z^{2}-\mu^{2}}-(-1)^{\sigma}\nu\right)F(z) \ , \quad  &
|z|>\mu
\end{array}
\right.
\label{Fbarpm}
\end{equation}
for a given function $F(z)$, and have taken into account that for
$z=iy$
\begin{equation}
(z^2+\mu ^2)^{1/2}=\left\{
\begin{array}{ll}
(\mu ^2-y^2)^{1/2} \ ,\quad  & |y|<\mu  \\
(y^2-\mu ^2)^{1/2}e^{i\pi /2} \ ,\quad  & y>\mu \\
(y^2-\mu ^2)^{1/2}e^{-i\pi /2} \ ,\quad  & y<-\mu
\end{array}
\right.   \label{sqroot}
\end{equation}
In this paper, we apply formula (\ref{sumJ1}) to the sums over $s$
in the expressions (\ref{qrin}) for the vacuum expectation values
of the energy density and vacuum stresses. As it can be seen from
expressions (\ref{fnueps})--(\ref{fnupperp}), the corresponding
functions $f(z)$ satisfy the relation
\begin{equation}
\label{relforf}
e^{-\nu \pi i}f(xe^{\pi i/2})=-e^{\nu \pi i}f(xe^{-\pi i/2}) \ ,
\quad {\mathrm{for}} \quad 0\leq x<\mu \ .
\end{equation}
By taking into account that for these $x$ one has $F^{(+)}=F^{(-)}$, we conclude
that the part of the integral on the right of Eq. (\ref{sumJ1}) over the interval
$(0,\mu )$ vanishes.

\end{document}